\documentclass[10pt]{article}

\usepackage{amsmath}

\usepackage{array}

\usepackage{appendix}

\usepackage{tocloft}                   % Adds Part structure to table of contents

\usepackage{graphicx}

\usepackage{amsfonts}

\usepackage{amssymb}

\usepackage{mathrsfs}

\usepackage{yfonts}

\usepackage{euscript}

\usepackage{centernot}                 % Permits me to trivially make a centred not version of any object

\usepackage{ifsym}                     % Gives a nicer sharp

\usepackage{upgreek}

\usepackage{mathtools}

\usepackage{bm}

\usepackage{color}

\usepackage{slantsc}
\usepackage{calligra}

\usepackage{bbold}          % Gives mathbb acting on lower case and, hopefully, numbers...

\usepackage[T1]{fontenc}

\usepackage{epsf}

\usepackage{latexsym}

\usepackage{tipa}

\usepackage{makeidx}

\makeindex

\def\be{\begin{equation}}

\def\ee{\end{equation}}

\def\beq{\begin{equation}}

\def\eeq{\end{equation}}

\def\bea{\begin{eqnarray}}

\def\eea{\end{eqnarray}}

\def\!{\hspace{-1.6667em}}

\def\m{\mbox{ }}

\def\mma {\m , \m \m }

\def\!{\hspace{-1.6667em}}

\def\n{\noindent}

\def\u{\underline}

\def\es{\m = \m}

\def\:={\m := \m}

\def\=:{\m =: \m}

\def\sbiQ{\mbox{\scriptsize\boldmath$Q$}}

\def\medwedge{\mbox{\large{$ \bm{\wedge}$}}}

\def\bcr{\mbox{$\,\, \bm{\times} \,\,$}}

\def\cr{\mbox{\scriptsize\bm{ $\mbox{ } \times \mbox{ }$}}}

\def\sumi2{\sum\mbox{}_{\mbox{}_{\mbox{\scriptsize $i$=1}}}^2}

\def\sumi3{\sum\mbox{}_{\mbox{}_{\mbox{\scriptsize $i$=1}}}^3}

\def\sumABcycles3{\sum\mbox{}_{\mbox{}_{\mbox{\scriptsize cycles $A,B$=1}}}^{3}}

\def\sumCDcycles3{\sum\mbox{}_{\mbox{}_{\mbox{\scriptsize cycles $C,D$=1}}}^{3}}

\def\sumj3{\sum\mbox{}_{\mbox{}_{\mbox{\scriptsize $j$=1}}}^3}

\def\sumk3{\sum\mbox{}_{\mbox{}_{\mbox{\scriptsize $k$=1}}}^3}

\def\prodiA1{\prod\mbox{}_{\mbox{}_{\mbox{\scriptsize $i$=1}}}^{A - 1}}

\def\d{\textrm{d}}                                                  % ordinary derivative

\def\pa{\partial}                                                   % partial derivative

\def\Hilb{\mbox{{\boldmath$\mathfrak{H}$}ilb}}                 % Hilbert space

\def\Phase{\mbox{{\boldmath$\mathfrak{P}$}hase}}                     % Phase space.

\def\bFrR{\mbox{\boldmath$\mathfrak{R}$}}                            % First letter of RigPhase, also used for Riem etc.  Is also, by itself, a ring.
\def\Rig-Phase{\bFrR\mbox{ig-}\Phase}                                % Rigged Phase Space

\def\Positive-Modespace{\mbox{{\boldmath$\mathfrak{M}$}odespace$^+$}}% Positive modespace

\def\POSITIVE-MODESPACE{\mbox{{\boldmath$\mathfrak{M}$}ODESPACE$^+$}}% Positive modespace alongside scalar field matter inhomogeneous modes.

\def\Kin-Hilb{\mbox{{\boldmath$\mathfrak{K}$}in-\Hilb}}                     % Dynamical Hilbert space 

\def\Mid-Hilb{\mbox{{\boldmath$\mathfrak{M}$}id-\Hilb}}                     % Dynamical Hilbert space 

\def\Dyn-Hilb{\mbox{{\boldmath$\mathfrak{D}$}yn-\Hilb}}                     % Dynamical Hilbert space 

\def\5Star{\mbox{\Large$\star$}}              % Rectified time derviative actually used

\begin{document}

\begin{titlepage}

\begin{center}

\vspace{0.1in}

\Large{\bf Specific PDEs for Preserved Quantities in Geometry. II.}

\vspace{0.1in}

\Large{\bf Affine Transformations and Subgroups.} \normalsize

\vspace{0.15in}

{\large \bf Edward Anderson$^*$}

\vspace{.15in}

\end{center}

\begin{abstract}
				
We extend finding geometrically-significant preserved quantities by solving specific PDEs to the affine transformations and subgroups.
This can be viewed not only as a purely geometrical problem but also as a subcase of finding physical observables, 
and furthermore as part of the comparative study of Background Independence level-by-level in mathematical structure. 
While cross and scalar-triple products (combined with differences and ratios) suffice to formulate these preserved quantities in 2- and 3-$d$ respectively,  
the arbitrary-dimensional generalization evokes the theory of forms.  
The affine preserved quantities are ratios of $d$-volume forms of differences, $d$-volume forms being the `top forms' supported by dimension $d$, 
and referring moreover to $d$-volumes of relationally-defined subsystems.  
								
\end{abstract}

\n Mathematics keywords: Geometrical automorphism groups and the corresponding preserved quantities.  
Geometrically-significant PDEs. Characteristic Problem. Shape Theory. Affine Geometry. 

\vspace{0.1in}
  
\n PACS: 04.20.Cv, 04.20.Fy, Physics keywords: observables, Background Independence. $N$-Body Problem. 

\vspace{0.1in}
  
\n $^*$ Dr.E.Anderson.Maths.Physics@protonmail.com

%==================================================================================================================================================================================
%==================================================================================================================================================================================
\section{Introduction}
%==================================================================================================================================================================================
%==================================================================================================================================================================================

\n We continue our program of geometrical preserved quantities (in Article I \cite{PE-1}'s sense) being 
systematically derived as solutions to PDE systems -- preserved equations -- treated as Free \cite{CH1} Characteristic Problems \cite{CH2, John}, 
\cite{PE-1} moreover providing specific methods of solution for these. 
%
%FFFFFFFFFFFFFFFFFFFFFFFFFFFFFFFFFFFFFFFFFFFFFFFFFFFFFFFFFFFFFFFFFFFFFFFFFFFFFFFFFFFFFFFFFFFFFFFFFFFFFFFFFFFFFFFFFFFFFFFFFFFFFFFFFFFFFFFFFFFFFFFFFFFFFFFFFFFFFFFFFFFFFFFFFFFFFFFFFFFFFFFFF
{            \begin{figure}[!ht]
\centering
\includegraphics[width=0.8\textwidth]{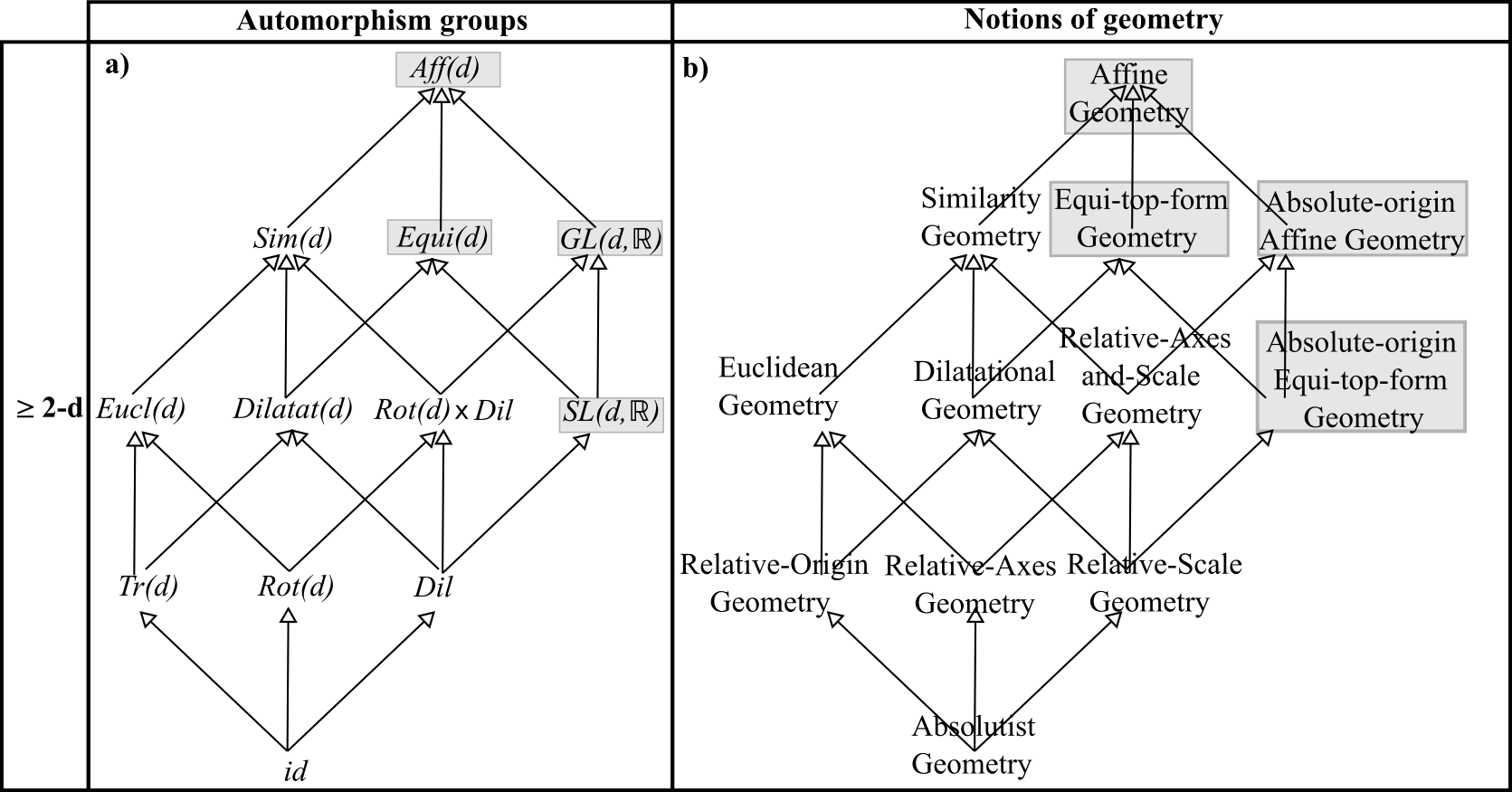}
\caption[Text der im Bilderverzeichnis auftaucht]{        \footnotesize{The current Article extends Article I by the cases highlighted in grey, 
to the lattice of geometrically-significant subgroups of the affine group a), and the corresponding geometries b).}}
\label{Aff-Latt} \end{figure}          }
%FFFFFFFFFFFFFFFFFFFFFFFFFFFFFFFFFFFFFFFFFFFFFFFFFFFFFFFFFFFFFFFFFFFFFFFFFFFFFFFFFFFFFFFFFFFFFFFFFFFFFFFFFFFFFFFFFFFFFFFFFFFFFFFFFFFFFFFFFFFFFFFFFFFFFFFFFFFFFFFFFFFFFFFFFFFFFFFFFFFFFFFFF

\m 

\n The current Article extends our repertoire of examples from Article I's Similarity Geometry to Affine Geometry \cite{Coxeter, Martin, Silvester}.
Each Article considers moreover the geometrically-significant continuous subgroups of the Article's `top geometry's' automorphism group: 
the similarity group $Sim(d)$ in Article I and the affine group $Aff(d)$ in the current Article. 
With 
\be 
Sim(d) \m < \m Aff(d)  \m , 
\ee 
what is covered by this extension is demarcated in grey in Fig \ref{Aff-Latt}. 

\m 

\n Geometrically significant subgroups of $Aff(d)$ which are not already subgroups of $Sim(d)$ are $SL(d, \,  \mathbb{R})$, $GL(d, \,  \mathbb{R})$ and $Equi(d)$. 
These are, respectively, the special-linear and general-linear groups over $\mathbb{R}$, and the equi-$d$-voluminal alias equi-top-form group. 
The last of these corresponds to Equi-$d$-voluminal alias Equi-top-form Geometry, which 

\end{titlepage}

\n requires a bit more explanation. 
In 2-$d$, this is somewhat better-known, under the name of Equiareal Geometry \cite{Coxeter}.  
Area is moreover 2-volume, rendering our extension to 3 and $d$ dimensions clear.  
`Top form' refers to the maximal $p$-form supported in dimension $d$.  

\m 

\n Ab initio, the form taken by Affine Geometry's preserved quantities is rather less obvious than Article 1's $Sim(d)$ examples, 
since Affine Geometry is usually set up to respect parallelism -- a relation -- rather than any target quantities.
This is also to be contrasted with Conformal Geometry (Article V), for which preservation of local-angle quantities is the most common starting point and motivation. 
Top forms is moreover furtherly relevant to how affine, $SL(d, \,  \mathbb{R})$, $GL(d, \,  \mathbb{R})$ preserved quantities are formulated.  
The Conclusion completes Fig 1 to Fig 3 with additional display of the dual bounded lattice of preserved quantities.   

\m

\n Preserved quantities as conceived of in the current Series of Articles are moreover underlied by consideration of constellations, constellation spaces, shapes and 
shape spaces \cite{Kendall84, Kendall89, Small, Kendall, FileR-Quad-I, PE16, ABook, I-II-III, Minimal-N}. 
In the context of Affine Geometry, the corresponding Affine Shape Theory has been developed and reviewed in particular in \cite{Sparr98, MP03, GT09, Bhatta, PE16}.   
Its main application to date is to Shape Statistics \cite{Kendall, JM00, Bhatta, DM16, PE16} in connection with Image Analysis \cite{Images} and Computer Vision \cite{CV}.  
Affine and Equi-top-voluminal Mechanics have moreover also recently been formulated in \cite{AMech} 
and used in a comparative analysis \cite{I89-I91, ASoS, ABook, Minimal-N-2} 
to all levels of structure of Background Independence \cite{BI, BI-2, Giu06, APoT, ABook, 5-6-7}.  
One of the nine conceptual aspects of Background Independence moreover concerns observables \cite{DiracObs, BI-2, ABeables}, 
for which Article I showed that the theory of geometrical preserved quantities is a subcase; 
affine observables are outlined in \cite{AObs2, AObs3}.

%==================================================================================================================================================================================
%==================================================================================================================================================================================
\section{$SL(2,\mathbb{R})$}
%==================================================================================================================================================================================
%==================================================================================================================================================================================

Let us first consider the 2 new preserved equations piecemeal. 
For $N = 1$, and setting $q_{1x} = x$ and $q_{1y} = y$, the first of these PDEs is  
\be 
( x \, \pa_x - y \, \pa_y )\sbiQ = 0  \m , 
\ee 
corresponding to {\it Procrustes stretches}.  

\m 

\n By the flow method, this is equivalent to the ODE system
\be
\dot{x} = x      \m ,
\ee
\be 
\dot{y} = -y     \m ,
\ee
\be 
\dot{\sbiQ} = 0  \m ,
\ee
to be solved as a Free Characteristic Problem. 
Integrating, 
\be 
x = u \, \mbox{exp}(t)  \m , 
\label{Int-1}
\ee
\be 
y = \mbox{exp}(t)       \m , 
\label{Int-2}
\ee
\be 
\sbiQ = \sbiQ(u)        \m .
\label{Int-3}
\ee 
Next, eliminating $t$ between (\ref{Int-1}) and (\ref{Int-2}) the characteristic coordinate takes the {\it product form}
\be
u = x \, y              \m .  
\ee 
Finally, substituting this in (\ref{Int-3}) the preserved quantities are
\be 
\sbiQ = \sbiQ(x \, y)   \m : 
\ee 
solution by suitably-smooth functions of products.

\m 

\n The second PDE is 
\be 
( x \, \pa_y + y \, \pa_x ) \sbiQ = 0  \m , 
\ee 
corresponding to {\it shears}.  

\m 

\n By the flow method, this is equivalent to the ODE system
\be
\dot{x} = y                 \m ,
\label{ODE-1}
\ee
\be 
\dot{y} = x                 \m ,
\label{ODE-2}
\ee
\be 
\dot{\sbiQ} = 0              \m ,
\label{ODE-3}
\ee
to be solved as a Free Characteristic Problem. 
So, differentiating (\ref{ODE-1}) and applying (\ref{ODE-2}), 
\be 
\ddot{x} = \dot{y} = x      \m , 
\ee  
by which  
\be 
x = u \, \mbox{cosh} \, t   \m , 
\label{int-4}
\ee
\be 
y = u \, \mbox{sinh} \, t   \m , 
\label{int-5}
\ee
\be 
\sbiQ = \sbiQ(u)              \m .
\label{int-6}
\ee 
Next, eliminating $t$ between (\ref{int-4}) and (\ref{int-5}), the characteristic coordinate takes the {\it difference of two squares form}
\be
u = x^2 - y^2               \m .
\ee 
Finally, substituting this in (\ref{int-6}) the preserved quantities are  
\be 
\sbiQ = \sbiQ(x^2 - y^2)      \m : 
\ee 
suitably-smooth functions of differences of squares.  

\m 

\n{\bf Remark 1} These individual results are noted side by side in \cite{CH2}. 

\m

\n{\bf Remark 2} $x = const$ maps to 
\be 
v  \es  - \frac{u^2}{x^2} + x^2  \m , 
\ee 
whereas $y = const$ maps to 
\be 
v  \es  \frac{u^2}{y^2} - y^2  \m . 
\ee
Both of these are families of confocal parabolas;  
\be 
\xi := u
\ee 
and 
\be
\eta \:=  \frac{v}{2}  \m , 
\ee 
are moreover orthogonal coordinates, constituting the standard 2-$d$ {\sl parabolic coordinates} \cite{MFI}.  

\m 

\n{\bf Remark 3} The corresponding change of coordinates furthermore decouples our preserved-equation PDEs to 
\be 
\pa_{\xi}\sbiQ = 0   \m ,
\ee 
\be 
\pa_{\eta}\sbiQ = 0  \m , 
\ee 
in each case provided that 
\be 
r = \sqrt{x^2 + y^2} \neq 0 \m .  
\label{Rad}
\ee 
This is to be compared with the rotation and dilation decoupling in Article I.  

\m 

\n{\bf Remark 4} This coordinate change decoupling does not however uplift to the $N$-point case, because each $N$ picks up its own radial factor (\ref{Rad}), 
resulting in these radial factors ceasing to be cancellable.  

\m 

\n{\bf Remark 5} A further complication is that the corresponding automorphism Lie algebra enforces 
that the shear and Procrustes preserved equations have to be considered alongside the rotational preserved equation. 
This means that $N = 2$ is minimal to have any nontrivial preserved quantities. 
Here, using $w = q_{1x}$, $x = q_{1y}$, $y = q_{2x}$, $z = q_{2y}$, our preserved equation system is 
\be 
( w \, \pa_x - x \, \pa_w + y \, \pa_z - z \, \pa_y ) \sbiQ  = 0                                                     \m , 
\ee 
\be 
( w \, \pa_x + x \, \pa_w + y \, \pa_z + z \, \pa_y ) \sbiQ  = 0  \m \mbox{ (shear preserved equation) }             \m ,
\ee 
\be 
( w \, \pa_w - x \, \pa_x + y \, \pa_y - z \, \pa_z ) \sbiQ  = 0  \m \mbox{ (2-$d$ Procrustes preserved equation) }  \m .  
\ee 
\n{\bf Remark 6} Solving the rotational preserved equation in the multipolar coordinates adapted to it individually does now not readily extend: 
the shear-and-Procrustes preserved equations are quite complicated in multipolars. 

\m 

\n{\bf Remark 7} The method which keeps matters simple is to consider the sum and the difference of the shear-and-rotation equations, 
\be 
( w \, \pa_x + y \, \pa_z ) \sbiQ  = 0  \m , 
\ee
\be 
( x \, \pa_w + z \, \pa_y ) \sbiQ  = 0  \m . 
\ee  
The first of these equations, treated piecemeal, is equivalent by the flow method to the ODE system 
\be 
\dot{x} = w    \m ,
\ee 
\be 
\dot{z} = y    \m ,
\ee 
\be 
\dot{\sbiQ} = 0  \m ,
\ee 
to be treated as a Free Characteristic Problem. 
Integrating, 
\be 
x = w \, t + u   \m ,
\label{int-1}
\ee 
\be 
z = y \, t     \m ,
\label{int-2}
\ee 
\be 
\sbiQ = \sbiQ(u, \, w, \, y)  \m .  
\label{int-3}
\ee 
Eliminating $t$ between (\ref{int-1}, \ref{int-2}), the form of the characteristics is 
\be 
u  \es  x - \frac{z}{y} w \m . 
\ee 
Finally, substtituting in (\ref{int-3}), 
\be 
\sbiQ = \sbiQ(w \, z - x \, y, \, w, \,  y)  \m . 
\ee 
A similar analysis of the second PDE, treated piecemeal, yields 
\be 
\sbiQ = \sbiQ(w \, z - x \, y, \, x, \, z)  \m . 
\ee 
Thus 
\be 
\sbiQ = \sbiQ(w \, z - x \, y)  \m . 
\ee
is the sole functional form of solution. 
While this has removed not two but three two functional dependencies, 
these two equations require the third -- Procrustes -- preserved equation to form a consistent system. 
Our solution moreover solves this as well, 
\be 
( w \, \pa_w - x \, \pa_x + y \, \pa_y - z \, \pa_z ) \sbiQ(w \, z - x \, y)  = \sbiQ^{\prime}(w \, z + x \, y - x \, y - w \, z) = 0  \m  
\ee 
for $^{\prime} = \d/\d A \mma A := w \, z - x \ y$. 
So, overall, our system of 3 equations uses up 3 functional dependencies, sending us from $2 N = 4$ freedoms to $4 - 3 = 1$ freedom.

\m 

\n{\bf Remark 8} 
\be 
w \, z - x \, y   \es  \left(\u{q}_1 \cr \u{q}_2\right)_{_{\mbox{\scriptsize $\perp$}}}  \es  Area(\u{q}_1, \u{q}_2)
\ee
turns out to be a lucid and generalizable interpretation of this preserved quantity.
This is the {\sl area} formed by two position vectors emanating from this model's absolute origin.
Due to their use of an absolute origin as one of their three defining points, we term these `absolute areas'.  

\m 

\n The above method furthermore generalizes to the general-$N$ case.  
The incipient system here is 

\n\be 
\sum_{I = 1}^N (x^I\pa_{y^I} - y^I\pa_{x^I})\sbiQ \es 0  \m , 
\ee

\n\be 
\sum_{I = 1}^N (x^I\pa_{x^I} + y^I\pa_{y^I})\sbiQ \es 0  \m , 
\ee

\n\be 
\sum_{I = 1}^N (x^I\pa_{x^I} - y^I\pa_{y^I})\sbiQ \es 0  \m . 
\ee
The sum and difference of the first two of these equations now gives 

\n\be 
\sum_{I = 1}^N x^I\pa_{y^I} \sbiQ \es 0  \m , 
\ee 

\n\be 
\sum_{I = 1}^N y^I\pa_{x^I} \sbiQ \es 0  \m .
\ee 
Solving these equations piecemeal gives, respectively,  
\be 
\sbiQ = \sbiQ(x^Ny^i - x^iy^N, \, x^I)  \m \mbox{ and } 
\ee 
\be 
\sbiQ = \sbiQ(x^Ny^i - x^iy^N, \, y^I)  \m . 
\ee
Thus 
\be 
\sbiQ = \sbiQ(x^Ny^i - x^iy^N)  \m ,  
\ee 
which moreover indeed solves the Procrustes preserved equation as well. 
Finally, using 
\be 
x^N y^i - x^i y^N   \es  \left(\u{q}^i \cr \u{q}^N\right)_{_{\mbox{\scriptsize $\perp$}}}  \es  Area(\u{q}^i, \, \u{q}^N) \m , 
\ee 
we summarize our result as the preserved quantities being of the form 
\be
\sbiQ = \sbiQ( \, {\bm{\cr}} \, )   \m :
\ee
suitably-smooth functions of the absolute areas.

\m 

\n{\bf Remark 9}  From the prespective of knowing that $SL(2, \mathbb{R})$ preserved quantities are areas, 
$N = 2$ is minimal to have nontrivial such, since these are absolute areas and it takes 2 distinct $\u{q}_I$ to form a such.

%==================================================================================================================================================================================
%==================================================================================================================================================================================
\section{$Equi(2)$}
%==================================================================================================================================================================================
%==================================================================================================================================================================================

\n The automorphism group for Equiareal Geometry is 
\be 
Equi(2)  \es  Tr(2) \rtimes SL(2, \mathbb{R})  \m .
\ee  
In this case, one has the 2 translational preserved equations alongside the previous section's triple, thus totalling 5 preserved equations.  
Article I's `centre of mass' sequential method moreover applies, 
sending one back to the previous section's triple of equations, for one object less and in terms of relative Jacobi coordinates $\rho_i$.  
Thus 
\be 
\sbiQ  \es  \sbiQ \left( \left(\u{\rho}^i \cr \u{\rho}^j\right)_{_{\mbox{\scriptsize $\perp$}}}\right)  
      \es  \sbiQ( \, {\bm{-\bcr-}} \, )                                                               \m :
\ee 
suitably-smooth functions of the {\sl areas} spanned by pairs of relative cluster vectors, now with no reference to any origin.

\m 

\n{\bf Remark 1} Defining an area requires 3 points. 
In the previous subsection, the absolute origin played the role of one of these points, 
whereas in the current subsection all three of the points are meaningfully realized, by which we term this notion of area `relational area'. 
It is moreover the area of a relationally-defined {\sl subsystem}.  
$N = 3$ is thus minimal to have nontrivial $Equi(2)$ preserved quantities, as this supports the minimum of $n = 2$ independent $\u{\rho}^i$ required to form a relational area.

%==================================================================================================================================================================================
%==================================================================================================================================================================================
\section{$GL(2,\mathbb{R})$}
%==================================================================================================================================================================================
%==================================================================================================================================================================================

One now has the preserved equation system 

\n\be 
\sum_{I = 1}^N x^I \pa_{y^I} \sbiQ  \es  0   \m ,  
\ee
which splits into the following quartet of preserved equations,

\n\be 
\sum_{I = 1}^N (x^I\pa_{y^I} - y^I\pa_{x^I})\sbiQ  \es  0  \m , 
\ee

\n\be 
\sum_{I = 1}^N (x^I\pa_{x^I} + y^I\pa_{y^I})\sbiQ  \es  0  \m , 
\ee

\n\be 
\sum_{I = 1}^N (x^I\pa_{x^I} - y^I\pa_{y^I})\sbiQ  \es  0  \m , 
\ee

\n\be 
\sum_{I = 1}^N (x^I\pa_{y^I} + y^I\pa_{x^I})\sbiQ  \es  0  \m . 
\ee 
We proceed by solving the $SL(2, \mathbb{R})$ system as per Sec 2 and the dilational preserved equation as per Sec I.7. 
This gives the {\it compatibility equation} 
\be 
\sbiQ( \, {\bm{ \cr}} \, )  \es  \sbiQ( \, {\bm{/}} \, )  \m , 
\ee 
which is solved by 
\be 
\sbiQ( \, {\bm{ \cr / \cr}} \, )                      \m :  
\ee 
suitably smooth functions of ratios of absolute areas.  

\m 

\n{\bf Remark 1} $N = 3$ is minimal to have nontrivial $GL(2, \mathbb{R})$ preserved quantities. 
For these are absolute area ratios, and we need 2 distinct $\u{q}^I$ for the first absolute area and at least one further $\u{q}^I$ for the second to be distinct, e.g.\ 
\be 
\left(\u{\rho}_1 \cr \u{\rho}_2\right)_{_{\mbox{\scriptsize $\perp$}}} \m \mbox{ and } \m  
\left( \u{\rho}_1 \cr \u{\rho}_3\right)_{_{\mbox{\scriptsize $\perp$}}}   \m . 
\ee

%==================================================================================================================================================================================
%==================================================================================================================================================================================
\section{$Aff(2)$}
%==================================================================================================================================================================================
%==================================================================================================================================================================================

The automorphism group for 2-$d$ Affine Geometry is 
\be 
Aff(2) \es  Tr(2) \rtimes GL(2, \mathbb{R})  \m .
\ee 
This case has the 2 translational preserved equations alongside the preceding section's quartet, thus totalling 6 preserved equations.  
The sequential method moreover applies, sending one back to the previous section's quartet, 
for one object less and in terms of relative Jacobi coordinates $\u{\rho}_i$.  
Thus 
\be 
\sbiQ = \sbiQ \left( \frac{ \left(\u{\rho}^i \cr \u{\rho}^j\right)_{_{\mbox{\scriptsize $\perp$}}}}{ \left(\u{\rho}^k \cr \u{\rho}^l\right)_{_{\mbox{\scriptsize $\perp$}}}}\right) 
\es \sbiQ( \, {\bm{\bcr/\bcr}} \, )  \m ,                                   
\ee 
suitably-smooth functions of ratios of relative areas.

\m 

\n{\bf Remark 1} This brings to the forefront that {\sl 2-d Affine Geometry preserves relational area ratios}. 
This is not obvious ab initio, since Affine Geometry is set up to preserve parallelism -- a relation rather than an object. 
A reasonably well-known first hint of involvement of area ratios can be found in Routh's Theorem \cite{Routh}. 
%
%FFFFFFFFFFFFFFFFFFFFFFFFFFFFFFFFFFFFFFFFFFFFFFFFFFFFFFFFFFFFFFFFFFFFFFFFFFFFFFFFFFFFFFFFFFFFFFFFFFFFFFFFFFFFFFFFFFFFFFFFFFFFFFFFFFFFFFFFFFFFFFFFFFFFFFFFFFFFFFFFFFFFFFFFFFFFFFFFFFFFFFFFF
{            \begin{figure}[!ht]
\centering
\includegraphics[width=0.6\textwidth]{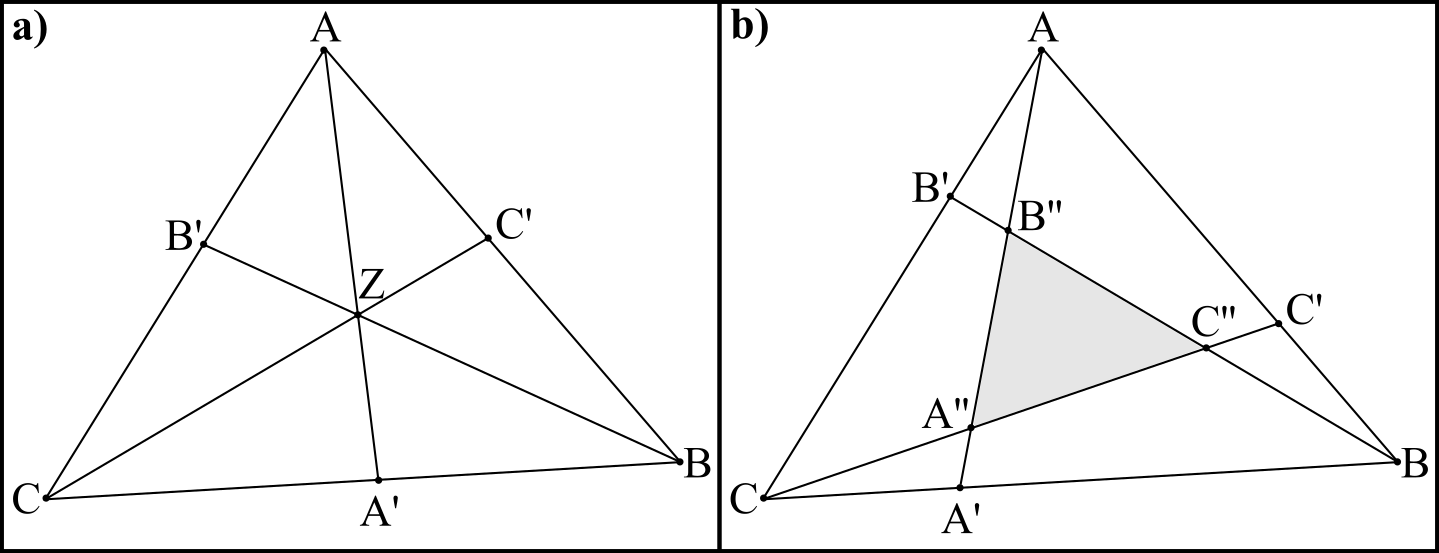}
\caption[Text der im Bilderverzeichnis auftaucht]{        \footnotesize{a) Cevians concurrent at some point $Z$. 
                                                                        b) Non-concurrent Cevians forming the Cevian triangle (shaded).}}
\label{Rout-Thm} \end{figure}          }
%FFFFFFFFFFFFFFFFFFFFFFFFFFFFFFFFFFFFFFFFFFFFFFFFFFFFFFFFFFFFFFFFFFFFFFFFFFFFFFFFFFFFFFFFFFFFFFFFFFFFFFFFFFFFFFFFFFFFFFFFFFFFFFFFFFFFFFFFFFFFFFFFFFFFFFFFFFFFFFFFFFFFFFFFFFFFFFFFFFFFFFFFF
	
\m 

\n Firstly, given a triangle, if a particular notion of Cevians for it are concurrent (Fig 2.a), 
\be 
a := \frac{CA^{\prime}}{BA^{\prime}} \m  \mbox{ and cycles }
\ee 
obey 
\be 
a \, b \, c = 1
\ee 
by Ceva's Theorem, a theorem which moreover has subsequently been found to have affine significance.  

\m 

\n Secondly, if a particular notion of Cevians for it are not concurrent, Routh found that 
\be 
\frac{  ( \mbox{signed Area of Cevian triangle } A^{\prime\prime}B^{\prime\prime}C^{\prime\prime} )  }{  (\mbox{Area of original triangle } ABC)  }  \es  
\frac{  (a \, b \, c - 1)^2  }{  ( 1 + a( 1 + c ) )( 1 + b ( 1 + a ) )(1 + c ( 1 + b ) )  }                                                                                   \m . 
\ee 
Note that the right hand side is evaluated using Menelaus' Theorem -- another affine-level theorem --  and is zero if the `concurrentor'
\be 
a \, b \, c - 1
\ee 
measuring departure from Ceva's concurrency is zero.  
Note finally that the left-hand-side of Routh's Theorem is an area ratio, i.e.\ a 2-$d$ affine invariant, thus constituting the aforementioned `hint'.  

\m

\n{\bf Remark 2}  $N = 4$ is minimal to have nontrivial $Aff(2)$ preserved quantities.
For these are relational area ratios and $N = 4$ supports $n = 3$ independent $\u{\rho}^i$, 
and we need 2 distinct $\u{\rho}^i$ to form a relational area, and at least 1 different $\u{\rho}^i$ to form a distinct second.

%==================================================================================================================================================================================
%==================================================================================================================================================================================
\section{$SL(3, \mathbb{R})$}
%==================================================================================================================================================================================
%==================================================================================================================================================================================

The preserved equations now form the system  

\n\be 
\sum_{I = 1}^N (x^{aI}\pa_{bI} - x^{bI}\pa_{aI})\sbiQ \es 0                   \m ,
\ee

\n\be 
\sum_{I = 1}^N (x^{aI}\pa_{bI} + x^{bI}\pa_{aI}) \sbiQ \es 0                  \m , 
\ee

\n\be 
\sum_{I = 1}^N (x^I\pa_{x^I} - y^{I}\pa_{y^I})\sbiQ \es 0                     \m , 
\ee

\n\be 
\sum_{I = 1}^N (x^{I}\pa_{x^I} y_I\pa_{y^I} - 2 \, z^{bI}\pa_{z^I})\sbiQ \es 0   \m .
\ee
There are 3 of the first type of equation, 3 of the second, 1 of the third and 1 of the fourth, totalling 8 preserved equations.  
Individually, only the last of these equations is of a new type: the {\it `hypercharge' preserved equation}.   
This is Particle Physics terminology \cite{Weinberg2}, based on $SL(3, \mathbb{R})$'s close analogy with $SU(3)$, which also in particular has a 
\be
\mbox{diag}(1, 1, -2)
\ee 
matrix in its $3 \times 3$ `Gell--Mann $\lambda$-matrix representation \cite{BK08}.  

\m 

\n The $N = 1$ case of our new equation is equivalent by the flow method to the ODE system
\be 
\dot{x} = x               \m ,
\ee 
\be 
\dot{y} = y               \m ,
\ee
\be 
\dot{z} = - 2 \, z        \m , 
\ee
\be 
\dot{\sbiQ} = 0            \m , 
\ee
to be solved as a Free Characteristic Problem. 
Integrating, 
\be 
x = \mbox{exp}(t)         \m , 
\label{int-7}
\ee
\be 
y = u \, \mbox{exp}(t)    \m , 
\label{int-8}
\ee
\be 
z = v  \, \mbox{exp}(-2 \, t)  \m ,
\label{int-9}
\ee
\be 
\sbiQ = \sbiQ(u, v)       \m . 
\label{int-10}
\ee 
Next,  eliminating $t$ by use of (\ref{int-7}) in (\ref{int-8}-\ref{int-9})
\be
u \es \frac{x}{y}         \m , 
\label{char-1}
\ee 
\be 
v = z \, x^2              \m . 
\label{char-2}
\ee
Finally, substituting (\ref{char-1}-\ref{char-2}) in (\ref{int-10}), 
\be 
\sbiQ \es \sbiQ \left(  \frac{x}{y} \mma  z \, x^2 \right)  \m . 
\ee 
Note that this dependence includes 
\be 
\frac{y}{x} \, z \, x^2  \es  x \, y \, z 
\ee 
dependence, as well as the $x$-$y$ symmetrizing 
\be 
\left( \frac{y}{x} \right)^2 \, z \, x^2  \es  z \, y^2    
\ee 
dependence. 

\m 

\n Solving the whole system, we find that the preserved quantities are 
\be 
\sbiQ  \es  \sbiQ( \, [\u{q}^I, \u{q}^J, \u{q}^K] \, ) 
      \es  \sbiQ( \, {\bm{[\m,\m,\m]}} \, )             \m :   
\ee
suitably-smooth functions of scalar triple products $[ \m , \m  , \m ]$. 
These can moreover be interpreted as the volumes made by triples of position vectors $\u{q}^I$, $\u{q}^J$, $\u{q}^K$, with $I$, $J$, $K$ all distinct. 
These vectors being with reference to the origin, this is an absolute notion of subsystem volume. 

\m 

\n{\bf Remark 1}  $N = 3$ is minimal as regards realizing nontrivial $SL(3, \mathbb{R})$ preserved quantities. 
For these are absolute subsystem volumes, and we need 3 distinct $\u{q}^I$ to form a such.

%==================================================================================================================================================================================
%==================================================================================================================================================================================
\section{$Equi(3)$}
%==================================================================================================================================================================================
%==================================================================================================================================================================================

\n The automorphism group for Equivoluminal Geometry is 
\be 
Equi(3)  \es  Tr(3) \rtimes SL(3, \mathbb{R})  \m .
\ee 
This case has 3 translational preserved equations alongside the preceding section's eightfold, thus totalling 11 preserved equations.   
The sequential method moreover applies, sending one back to the previous triple of equations, for one object less and in relative Jacobi coordinates $\rho^i$.  
Thus 
\be 
\sbiQ  \es  \sbiQ( \, [\u{\rho}^i, \, \u{\rho}^j, \, \u{\rho}^k] \, )  
      \es  \sbiQ( \, {\bm{[- , - , -]}} \, )                                \m :
\ee 
suitably-smooth functions of the {\sl volumes} spanned by triples of relative cluster vectors, now with no reference to any absolute origin.

\m 

\n{\bf Remark 1} Defining a volume requires 4 points.
In the previous subsection, the absolute origin was playing the role of one of these points, 
whereas in the current subsection all four of the points are meaningful rather than absolute, by which we term this subsystem notion of volume `relational volume'. 

\m 

\n{\bf Remark 2}  $N = 4$ is minimal to have nontrivial $Equi(3)$ preserved quantities.
For these are relational volumes, and $N = 4$ supports $n = 3$ independent $\u{\rho}^i$, and we need 3 distinct $\u{\rho}^i$ to form a relational volume.

%==================================================================================================================================================================================
%==================================================================================================================================================================================
\section{$GL(3, \mathbb{R})$}
%==================================================================================================================================================================================
%==================================================================================================================================================================================

The preserved equations are now 

\n\be 
\sum_{I = 1}^N x^{aI} \pa_{bI} \sbiQ  \es  0                 \m ,  
\ee
which split into a quintet of types of preserved equations,

\n\be 
\sum_{I = 1}^N ( x^I \pa_{x^I} + y^I \pa_{y^I} + z^I \pa_{z^I} )\sbiQ \es 0  \m , 
\ee

\n\be 
\sum_{I = 1}^N ( x^{aI} \pa_{bI} - x^{bI} \pa_{aI} ) \sbiQ \es 0                     \m ,
\ee

\n\be 
\sum_{I = 1}^N ( x^{aI} \pa_{bI} + x^{bI} \pa_{aI} ) \sbiQ \es 0                     \m , 
\ee

\n\be 
\sum_{I = 1}^N ( x^{I}\pa_{x^I} - y^I\pa_{y^I} )\sbiQ \es 0                       \m , 
\ee

\n\be 
\sum_{I = 1}^N (x^{I}\pa_{x^I} + y^I\pa_{y^I} - 2 \, z^{I}\pa_{z^I})\sbiQ \es 0   \m .
\ee
There are 3 equations of the second kind, 3 of the third kind and 1 of each of the other kinds, thus totalling 9 preserved equations.

\m 

\n Using Secs 6 and I.7, we arrive at the compatibility equation 
\be 
\sbiQ( \, {\bm{/}} \, )  \es  \sbiQ( {\bm{ \, [\m,\m,\m]}} \, )                                            \m , 
\ee 
which is solved by 
\be 
\sbiQ  \es  \sbiQ( \, {\bm{ [\m,\m,\m] \, / \, [\m,\m,\m]}} \, )                                                \m :
\ee 
suitably-smooth functions of ratios of absolute volumes. 

\m 

\n{\bf Remark 1}  $N = 4$ is minimal to have nontrivial $GL(3, \mathbb{R})$ preserved quantities. 
For these are absolute volume ratios, and we need 3 distinct $\u{q}^I$ to form an absolute volume, 
and at least one different $\u{q}^I$ to form a distinct second.

%==================================================================================================================================================================================
%==================================================================================================================================================================================
\section{$Aff(3)$}
%==================================================================================================================================================================================
%==================================================================================================================================================================================

\n The automorphism group for 3-$d$ Affine Geometry is 
\be 
Aff(3)  \es  Tr(3) \rtimes GL(2, \mathbb{R})  \m .
\ee 
This case has 3 translational preserved equation alongside the preceding section's 9, thus totalling 12 preserved equations.  
The sequential method moreover applies, sending one back to the previous quartet of equations, for one object less and in relative Jacobi coordinates $\rho_i$.  
Thus 
\be 
\sbiQ  \es  \sbiQ( \, {\bm{ [-,-,-] \, / \, [-,-,-]}} \, )                                      \m :
\ee 
suitably-smooth functions of the {\sl volumes} spanned by triples of relative cluster vectors, now with no reference to any absolute origin.

\m 

\n{\bf Remark 1}  $N = 5$ is minimal to have nontrivial $Aff(3)$ preserved quantities. 
For these are relational volume ratios, $N = 5$ supports $n = 4$ independent $\u{\rho}^i$, 
and we need 3 distinct $\u{\rho}^i$ to form a relational volume, and at least one different $\u{\rho}^i$ to form a distinct second, e.g.\ 
\be
[\u{\rho}_1, \, \u{\rho}_2, \, \u{\rho}_3]  \m \mbox{ and } \m
[\u{\rho}_1, \, \u{\rho}_2, \, \u{\rho}_4]  \m .
\ee

%==================================================================================================================================================================================
%==================================================================================================================================================================================
\section{$SL(d, \,  \mathbb{R})$}
%==================================================================================================================================================================================
%==================================================================================================================================================================================

We now have $d^2 - 1$ preserved equations. 
Relative to dimension $d - 1$, the only new type of equation in the preserved system corresponds to one of the Procrustes stretches,  

\n\be 
\sum_{I = 1}^N \left\{  \sum_{\bar{a} = 1}^{d - 1} x^{\bar{a}I} \pa_{x^{\bar{a}I}} - (d - 1) x^{dI} \pa_{x^{dI}}  \right\} \sbiQ  \es  0   \m .
\ee
The barred indices here run from 1 to $d - 1$.
We term this the {\it generalized `hypercharge' preserved equation}, since the $SL(3, \mathbb{R})$ to $SU(3)$ analogy extends to arbitrary $d$, 
as does the Gell--Mann matrix repesentation of the latter \cite{BK08}, now to  $d \times d$ matrices including in particular the matrix  
\be 
\mbox{diag}(1, \, 1, \, ... , \, 1, \, 1 - d)  
\ee 
of $SU(d)$'s `last generalized hypercharge'.

\m 

\n For $N = 1$, our new equation is  
\be 
\left\{   \sum_{\bar{a} = 1}^{d - 1} x^{\bar{a}}  \pa_{x_{\bar{a}}}   -  (d - 1)  x_{d}  \pa_{x_d}  \right\} \sbiQ  \es  0   \m .
\ee
This is equivalent by the flow method to the ODE system
\be 
\dot{x}^{\bar{a}} = x^{\bar{a}}                             \m , 
\ee 
\be 
\dot{x}_d = (1 - d) \, x_d                                  \m ,
\ee
\be 
\dot{\sbiQ} = 0                                              \m , 
\ee
to be solved as a Free Characteristic Problem. 
So integrating and using tilded indices to run from 1 to $d - 2$, 
\be 
x^{\widetilde{a}} = u^{\widetilde{a}}\mbox{exp}(t)            \m , 
\label{int-11}
\ee
\be 
x_{d - 1}  =  \mbox{exp}(t)                                   \m , 
\label{int-12}
\ee
\be 
x_d        =  v \, \mbox{exp}((1 - d)) \, t)                  \m , 
\label{int-13}
\ee
\be 
\sbiQ       =  \sbiQ(u^{\widetilde{a}}, \, v)                      \m . 
\label{int-14}
\ee 
Thus eliminating $t$ by use of (\ref{int-12}) in (\ref{int-11}, \ref{int-13}), we obtain the characteristic coordinates 
\be
u^{\widetilde{a}}  \es  \frac{x^{\widetilde{a}}}{x_{d - 1}}   \m , 
\label{char-3}
\ee 
\be 
v = x_d \, {x_{d - 1}}^{d - 1}                                \m ,
\label{char-4}
\ee
the last item's upper $d - 1$ indicating a power while the lower $d - 1$ is, as usual, an index. 
Finally, substituting (\ref{char-3}-\ref{char-4}) in (\ref{int-14}), we obtain the preserved quantities  
\be 
\sbiQ  \es  \sbiQ  \left(  \frac{  x^{\widetilde{a}}  }{  x_{d - 1}  } \mma  x_{d} \, {  x_{d - 1}  }^{d - 1} \right)  \m . 
\ee 
This dependence includes 
\be 
\prod_{\widetilde{a} = 1}^{d -2} \left(  \frac{x^{\widetilde{a}}}{x_{d - 1}}  \right)  x_{d} \, {x_{d - 1}}^{d - 1}  \es 
\left(\prod_{\widetilde{a} = 1}^{d -2} x^{\widetilde{a}} \right) x_{d - 1} x_d                                       \es  \prod_{a = 1}^d x^a  \m .
\ee 
\n Solving the whole system, we find 
\be 
\sbiQ  \es  \sbiQ \left( \bigwedge_{a = 1}^d \u{q}^{I_a} \right)  \m :
\ee
suitably-smooth functions of the $\bigwedge_{a = 1}^d$: $d$-volume forms; this being on $d$-dimensional space, it is moreover the {\sl top} form supported. 
This is to be interpreted as $d$-Volume, 
\be 
\mbox{$d$-Volume} = \bigwedge         \m ,
\ee 
formed by $d$ distinct-index position vectors relative to the absolute origin, in which sense it is an absolute $d$-volume concept.
This clearly incorporates 
\be 
\mbox{Area} = {\bm{ \bcr}}
\ee 
and 
\be 
\mbox{Volume} = {\bm{  [\m,\m,\m]  }}
\ee
as the $d = 2$ and $3$ cases respectively.
Due to this, we supplant the use of ${\bm{\bcr}}$ and ${\bm{ [\m,\m,\m]}}$ shorthands by ${\bm{\medwedge}}$, denoting the current subsection's answer by 
\be 
\sbiQ \es \sbiQ(\, \medwedge \,)            \m : 
\ee 
suitably-smooth functions of the absolute top forms.  

\m 

\n{\bf Remark 1} $N = d$ is minimal as regards realizing any $SL(d, \,  \mathbb{R})$ preserved quantities. 
For these are absolute subsystem $d$-volumes, and $d$ quantities $\u{q}_I$ are required to realize one of these.  
We comment further on the significance of Secs 10-13's minimal $N$'s in the Conclusion.

%==================================================================================================================================================================================
%==================================================================================================================================================================================
\section{$Equi(d)$}
%==================================================================================================================================================================================
%==================================================================================================================================================================================

The automorphism group for Equi-$d$-voluminal alias Equi-top-form Geometry is 
\be 
Equi(d)  \es  Tr(d) \rtimes SL(d, \,  \mathbb{R})  \m .
\ee 
This case has $d$ translational preserved equations alongside the preceding section's system of $d^2 - 1$ equations, thus totalling $d^2 + d - 1$ preserved equations.  
The sequential method moreover applies, sending one back to the above system of equations, for one object less and in relative Jacobi coordinates $\rho_i$.  
Using also that ${\bm{\medwedge}}$ is $d$-nary, unlike ${\bm{-}}$, ${\bm{/}}$ ${\bm{\cdot}}$ or ${\bm{\cr}}$, we thus have 
\be 
\sbiQ   \es  \sbiQ \left( \bigwedge_{a = 1}^d \u{\rho}^{i_a} \right)     
       \es  \sbiQ( \, {\bm{ \medwedge(-)}} \, )                                                            \m ,
\ee 
suitably-smooth functions of the {\sl d-volumes} alias {\it top forms} spanned by $d$-tuples of relative cluster vectors, now with no reference to any absolute origin.

\m 

\n{\bf Remark 1} Defining a $d$-volume requires $d + 1$ points. 
On the one hand, in the previous subsection, the absolute origin was playing the role of one of these points.  
On the other hand, in the current subsection all $d + 1$ of these points are meaningfully realized.  

\m 

\n{\bf Remark 2} Thus $N = d + 1$ is minimal to realize nontrivial $Equi(d)$ preserved quantities.
For these are relationally significant volumes, $N = d + 1$ supports $n = d$ independent $\u{\rho}^i$, and $d$ distinct such are required to form a relational $d$-volume.

%==================================================================================================================================================================================
%==================================================================================================================================================================================
\section{$GL(d, \,  \mathbb{R})$}
%==================================================================================================================================================================================
%==================================================================================================================================================================================

In this case, we have the $d^2$ preserved equations

\n\be 
\sum_{I = 1}^N x^{aI} \pa_{x^{bI}} \sbiQ  \es  0 \m . 
\ee
which split into the $d^2 - 1$ equations of Sec 10 alongside the dilational preserved equation 

\n\be 
\sum_{I = 1}^N (x^{aI}\pa_{x^{aI}} + y^{aI}\pa_{y^{aI}})\sbiQ  \es  0                   \m , 
\ee
Using Sec 10's result alongside the solution of the Euler homogeneity equation of degree 0, we arrive at the compatibility equation 
\be 
\sbiQ({\bm{/}})  \es  \sbiQ( \, {\bm{\medwedge}} \, )                                                \m , 
\ee 
which is realized by 
\be 
\sbiQ \es \sbiQ( \, {\bm{\medwedge/\medwedge}} \, )                                                \m :
\ee 
suitably-smooth functions of ratios of absolute volumes. 

\m 

\n{\bf Remark 1} $N = d + 1$ is minimal to realize nontrivial $GL(d, \,  \mathbb{R})$ preserved quantities. 
For these are absolute $d$-volume ratios, and we need $d$ distinct $\u{q}^I$ to realize 1 volume and the first, and at least 1 different $\u{q}^I$ to realize a distinct second.

%==================================================================================================================================================================================
%==================================================================================================================================================================================
\section{$Aff(d)$}
%==================================================================================================================================================================================
%==================================================================================================================================================================================

The automorphism group for $d$-dimensional Affine Geometry is 
\be 
Aff(d) \es  Tr(d) \rtimes GL(d, \,  \mathbb{R})  \m .
\ee 
This case has $d$ translational preserved equations alongside the preceding section's system of $d^2$ equations, thus totalling $d(d + 1)$ preserved equations.  
\n The $Aff(d)$ case has the translational preserved equation alongside the above system. 
But the sequential method applies, sending one back to the previous system of equations, for one object less and in relative Jacobi coordinates $\rho_i$.  

\m 

\n Thus 
\be 
\sbiQ  \es  \sbiQ( \, {\bm{ \medwedge(-) \, / \, \medwedge(-) }} \, )                                      \m : 
\ee 
suitably-smooth functions of the top forms spanned by $d$-tuples of relative cluster vectors, i.e.\ now with no reference to any absolute origin.

\m 

\n{\bf Remark 1} $N = d + 2$ is minimal to realize nontrivial preserved quantities in $d$-dimensional Affine Geometry. 
For these are relational $d$-volume ratios, and $N = d + 2$ supports $n = d + 1$ independent $\u{\rho}_i$, 
which is the minimum amount to form 2 distinct $d$-volume top forms, e.g.\ 
\be
\u{\rho}_1 \wedge \u{\rho}_2 \wedge ... \wedge \u{\rho}_{d - 1} \wedge \u{\rho}_d        \m \m \mbox{ and } \m \m  
\u{\rho}_1 \wedge \u{\rho}_2 \wedge ... \wedge \u{\rho}_{d - 1} \wedge \u{\rho}_{d + 1}  \m .  
\ee

%==================================================================================================================================================================================
%==================================================================================================================================================================================
\section{Conclusion}
%==================================================================================================================================================================================
%==================================================================================================================================================================================

The current article considers Affine transformations $Aff(d)$ and subgroups' systems of preserved-equation PDEs; 
these are larger than Article I's similarity transformations $Sim(d)$ counterparts.
The new equations restrict preserved quantities with respect to shears and Procrustes stretches; these both require dimension $d \geq 2$ to occur at all.
The shear preserved equations take the same form dimension by dimension; there are $d(d - 1)/2$ of them in dimension $d$, 
with each placing a difference-of-squares functional form restriction on the preserved quantities. 
Each dimension $d$ has one new type of Procrustes-stretch preserved equation relative to the previous dimension, for a total of $d - 1$ equations.
The 2-$d$ Procrustes equation places a product restriction on the preserved quantities, which generalizes to an
$x y^{m}$ functional form restriction for $m = 1$ to $d - 1$. 
The `top' Procrustes equation in 3-$d$ is closely related to the $SU(3)$ hypercharge of Particle Physics, with subsequent dimensions' further Procrustes equations 
likewise bearing close relation to $SU(d)$'s generalized hypercharges. 
Both are underlied by the same kind of choice of basis of tracefree quantities \cite{BK08}: 
\be 
(1, \, -1, \, 0, \, ... \, , \, 0) \mma (1, \, 1, \, -2, \, 0, \, ... \, , \, 0) \mma ... (1, \, 1, \, ..., 1, \, 1 - d)  \m . 
\ee 
Due to the brackets formed by the underlying generators, shears and Procrustes stretches moreover must occur both together and furthermore alongside rotation preserved equations.

\m 

\n Aside from the affine group itself, this leaves us with 3 other geometrically-significant groups aside those of $Sim(d)$ that were already covered in Article 1. 
Namely the real-valued general-linear group $GL(d, \,  \mathbb{R})$ and special-linear group $SL(d, \,  \mathbb{R})$ and the equi-top-form group $Equi(d)$.  
The corresponding notions of geometry are, respectively, Absolute-origin Affine Geometry, Absolute-origin Equi-top-form Geometry and Equi-top-form Geometry. 
We now extend the Introduction's Fig 1 -- of the bounded lattices of automorphism groups and of the corresponding Geometries -- 
to include also the dual bounded lattice of notions of preserved quantities as evaluated in the current Article by systematic solution of preserved equations, in Fig 3. 

\m 

\n In particular, firstly $Aff(d)$ has now supplanted $Sim(d)$ as `top group', i.e.\ the group that all the other geometrical automorphism groups considered are subgroups of. 
Secondly, ratios of top forms of differences have supplanted ratios of dot products of differences as the dual lattice's bottom -- most restricted -- preserved quantities. 
That affine invariants take this form is not a priori obvious, though, as we explained, Routh's Theorem \cite{Routh} offered an early hint of this in 2-$d$.  

\m

\n Our presentation of the dual bounded lattice of preserved quantities continues to benefit from Article I's compact notation ${\bm{\sim}}$, ${\bm{-}}$, ${\bm{/}}$, ${\bm{\cdot}}$,  
now with the extension to include a ${\bm{\medwedge}}$ symbol as well. 
In combinations, this features furthermore as ${\bm{\wedge}}(\m)$, since it, as the top form supported in dimension $d$ -- a $d$-volume form -- is $d$-nary, 
unlike the previous three symbols in the notation, which are all binary.  
A stepping stone in this regard used in the current Article is that the 2-$d$ version of ${\bm{\medwedge}}$ 
-- the 3-component of the cross product ${\bm{\cr}}$ for area = 2-volume forms -- is itself binary.  
We then solved in 3-$d$ in terms of the scalar triple product before bringing in the $d$-nary top form for the general case.  
%
%FFFFFFFFFFFFFFFFFFFFFFFFFFFFFFFFFFFFFFFFFFFFFFFFFFFFFFFFFFFFFFFFFFFFFFFFFFFFFFFFFFFFFFFFFFFFFFFFFFFFFFFFFFFFFFFFFFFFFFFFFFFFFFFFFFFFFFFFFFFFFFFFFFFFFFFFFFFFFFFFFFFFFFFFFFFFFFFFFFFFFFFFF
{            \begin{figure}[!ht]
\centering
\includegraphics[width=1.0\textwidth]{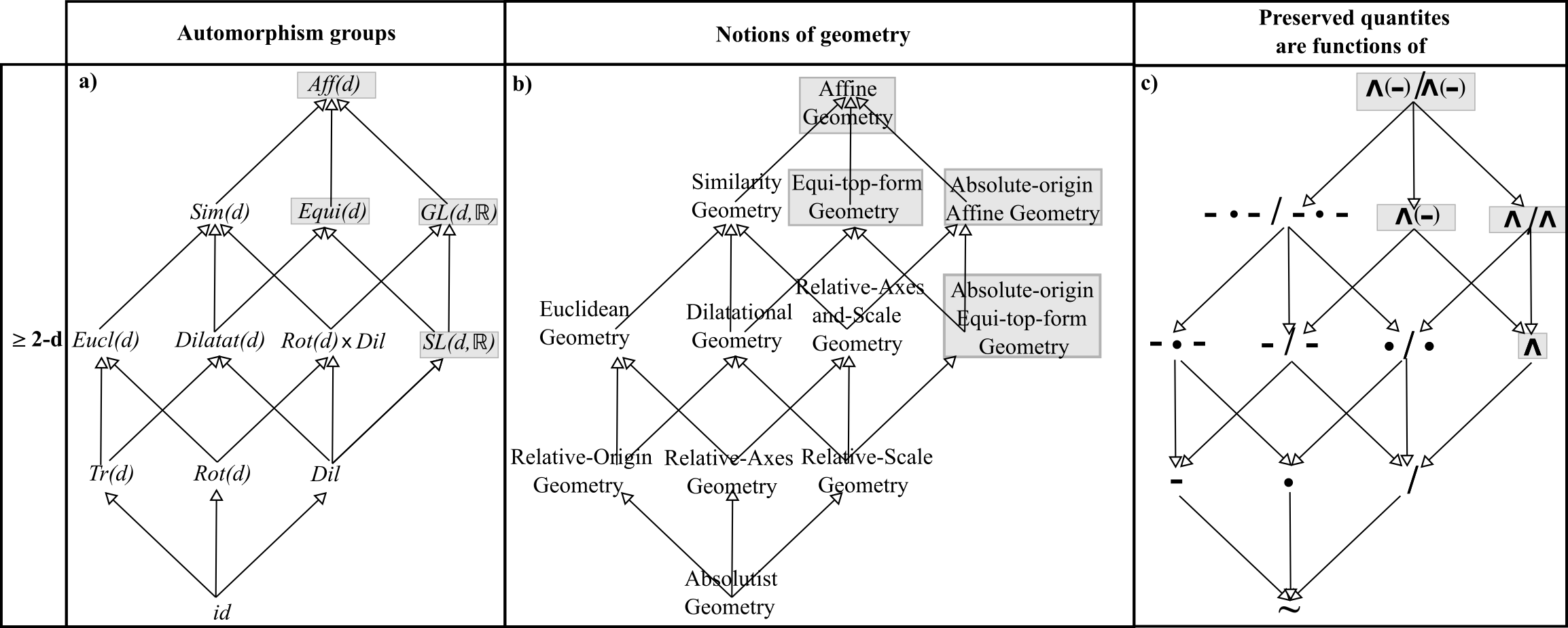}
\caption[Text der im Bilderverzeichnis auftaucht]{        \footnotesize{We here extend Fig 1 to include the dual bounded lattice of preserved quantities, 
comprising of suitably smooth functions of the indicated functional dependencies.}}
\label{Aff-Latt-2} \end{figure}          }
%FFFFFFFFFFFFFFFFFFFFFFFFFFFFFFFFFFFFFFFFFFFFFFFFFFFFFFFFFFFFFFFFFFFFFFFFFFFFFFFFFFFFFFFFFFFFFFFFFFFFFFFFFFFFFFFFFFFFFFFFFFFFFFFFFFFFFFFFFFFFFFFFFFFFFFFFFFFFFFFFFFFFFFFFFFFFFFFFFFFFFFFFF
	
\m 

\n In solving for the current Article's four further types of preserved quantity, Article I's sequential method resting on passing to centre of mass frame remains useful, 
reducing our working to two pairs of nested workings. 
The `compatibility equation' method is moreover useful in combining the resolution of the $SL(d, \,  \mathbb{R})$ preserved equation system 
by the top form with the remaining preserved equation of $GL(d, \,  \mathbb{R})$: the dilational equation of Euler homogeneity of degree zero type as solved by ratios. 
Between these two workings, we need only solve a single new system to get all four of the current Article's four geometrically-significant function spaces of preserved quantities. 

\m 

\n Two further research direction following from the current Article are as follows. 

\m 

\n{\bf Frontier 1} The $d$-dimensional equivoluminal model being nontrivial requires it to support at least one relational $d$-volume, requiring $N \geq d + 1$ points. 
$N = d + 1$  moreover characterizes the simplex, and is the case for which we have a basis of relative vectors: the so-called Casson diagonal, which is of further 
shape-theoretic significance as explained in \cite{Kendall, Minimal-N, A-Coolidge}.
The same relation $N = d + 1$ also applies to $GL(d, \,  \mathbb{R})$ models, since this is to support at least 2 distinct absolute $d$-volumes. 
On the other hand, a $SL(d, \,  \mathbb{R})$ model being nontrivial requires it to support at least 1 absolute $d$-volume, requiring $N \geq d$ points. 
These values form the first sub-Casson diagonal, i.e.\ the minimal nonspanning case \cite{Minimal-N}.
Finally, a $d$-dimensional affine model being nontrivial requires it to support at least one relational $d$-volume ratio, requiring $N \geq d + 2$ points.    
These values form the first super-Casson diagonal, i.e.\ the minimal linearly dependent case \cite{Minimal-N}.
For the familiar case of $d = 3$, the abover three values are 3, 4 and 5, thus forming the famous series of increasing complexities in passing from 3-Body Problems 
to 4-Body Problems and then to 5-Body Problems \cite{LR95-97, M02-M05-M15, Minimal-N}.  
The current Article thus provides a further structural piece of motivation for the 3-$d$ 5-Body Problem, as having the minimal $N$ to support having nontrivial affine content. 
This will moreover play a part in the comparative study of Background Independence, level by level in mathematical structure. 

\m 

\n\n{\bf Frontier 2} The current Article's top-form-ratio invariants enter the projective invariants of Article IV. 
This is in the context of the Affine Group itself being subgroup of the projective group, 
the further preserved equation in this case concurrently imposing, for $d \geq 2$  that the preserved quantities be concurrently both top-form-ratios and cross-ratios: 
yet another compatibility equation.  
 
\m 

\n{\bf Acknowledgments} I thank Chris Isham and Don Page for previous discussions.  
Enrique Alvarez, Jeremy Butterfield, Malcolm MacCallum and Reza Tavakol for support with my career.

%==================================================================================================================================================================================
%======================================================================  BIBLIOGRAPHY  ============================================================================================

\end{document}